\documentclass[%
 aip,
 amsmath,amssymb,
 reprint,%
]{revtex4-1}

\usepackage{graphicx}
\usepackage{dcolumn}
\usepackage{bm}

\usepackage[utf8]{inputenc}
\usepackage[T1]{fontenc}
\usepackage{mathptmx}
\usepackage{etoolbox}
\usepackage{subcaption}
\makeatletter
\def\@email#1#2{%
 \endgroup
 \patchcmd{\titleblock@produce}
  {\frontmatter@RRAPformat}
  {\frontmatter@RRAPformat{\produce@RRAP{*#1\href{mailto:#2}{#2}}}\frontmatter@RRAPformat}
  {}{}
}%
\makeatother
\begin{document}

\preprint{AIP/123-QED}

\title[Heavy-ion-driven PWFA with drift-like phase-shift sections]{Heavy ion driven plasma wakefield acceleration with drift-like phase-shift sections}
\author{Jiangdong Li}
 \affiliation{State Key Laboratory of Nuclear Physics and Technology, and Key Laboratory of HEDP of the Ministry of Education, CAPT, School of Physics, Peking University, Beijing 100871, China.}
\author{Jiancheng Yang}%
 \affiliation{Institute of Modern Physics, Chinese Academy of Sciences, Lanzhou 730000, China.}
 \affiliation{University of Chinese Academy of Sciences, Beijing 100049, China.}
\author{Guoxing Xia}
 \affiliation{University of Manchester, Manchester M13 9PL, United Kingdom.}
 \affiliation{Cockcroft Institute, Daresbury, Cheshire WA4 4AD, United Kingdom.}
\author{Guangxian Li}
 \affiliation{Institute of Modern Physics, Chinese Academy of Sciences, Lanzhou 730000, China.}
 \affiliation{University of Chinese Academy of Sciences, Beijing 100049, China.}
\author{Xueqing Yan}
 \affiliation{State Key Laboratory of Nuclear Physics and Technology, and Key Laboratory of HEDP of the Ministry of Education, CAPT, School of Physics, Peking University, Beijing 100871, China.}
\author{Jie Liu}
 \email{liujie115@impcas.ac.cn}
 \affiliation{Institute of Modern Physics, Chinese Academy of Sciences, Lanzhou 730000, China.}
 \affiliation{University of Chinese Academy of Sciences, Beijing 100049, China.}
\date{\today}

\begin{abstract}
In heavy-ion-driven plasma wakefield acceleration, ultra-relativistic witness electrons eventually outrun the accelerating phase of the wake excited by a subluminal driver. We investigate a drift-like phase-shift scheme that inserts periodic vacuum gaps between uniform-density plasma sections. Across each vacuum gap, the plasma wake vanishes and the witness advances relative to the heavy-ion driver, bypassing the decelerating and defocusing phases before re-entering the accelerating bucket in the next plasma stage. Using driver parameters relevant to the High Intensity Heavy-ion Accelerator Facility (HIAF), quasi-static LCODE simulations with idealized transverse confinement show that a $16\,\mathrm{MeV}$ witness electron bunch is accelerated to a peak energy of $700\,\mathrm{MeV}$ over $1.65\,\mathrm{m}$ with a relative energy spread of ${\sim}5\%$ while maintaining the wakefield amplitude. To test transport across physical boundaries, two-dimensional electromagnetic particle-in-cell simulations in EPOCH model the first plasma--vacuum--plasma unit with a finite-length alternating-gradient quadrupole lattice ($\le 600\,\mathrm{T/m}$). With prescribed initial beam convergence, the transverse rms width of the trailing tenth $\mathrm{Bi}$ microbunch broadens from $100$ to $104.5\,\mu\mathrm{m}$, and the passive witness electrons reach a mean kinetic energy of $141.1\,\mathrm{MeV}$ with an energy spread of $1.49\%$. When initial convergence is removed for both beams, the tenth $\mathrm{Bi}$ microbunch broadens to $139.1\,\mu\mathrm{m}$, yet the passive witness still reaches $120.9\,\mathrm{MeV}$ with a $1.57\%$ energy spread and $99.9\%$ retention. With the converging baseline entrance conditions and an equivalent witness charge of $0.3\,\mathrm{pC}$, the witness reaches a mean energy of $137.1\,\mathrm{MeV}$ and a $3.10\%$ energy spread with nearly complete transmission. These results verify first-unit transport, wake recovery, and reacceleration under specified entrance conditions, providing a numerical basis for future experimental design.
\end{abstract}

\maketitle

\section{\label{sec:intro}Introduction}

\subsection{\label{sec:level11}Plasma based acceleration}

High-energy particle accelerators are essential tools for probing the fundamental constituents of matter and their interactions. Currently, the Large Hadron Collider (LHC) at CERN delivers proton-proton collisions at a center-of-mass energy of $13.6\,\mathrm{TeV}$ in a $27\,\mathrm{ km}$ ring. To reach higher energies, CERN's proposed Future Circular Collider (FCC) \cite{FCC:2018byv} targets $100\,\mathrm{TeV}$ center-of-mass collisions in a $91\,\mathrm{km}$ tunnel, requiring tens of billions of euros and several decades of R\&D and construction across its FCC-ee \cite{Benedikt:2025kwi} and FCC-hh \cite{Benedikt:2025opw} stages. These scale and cost constraints motivate compact acceleration concepts, making plasma-based acceleration an attractive alternative.

Plasma-based acceleration, first proposed by Tajima and Dawson \cite{Tajima:1979bn}, uses intense laser pulses \cite{Esarey:2009zz,Pukhov:2002otp,Lu:2006nz} or charged-particle bunches \cite{Chen:1984up,Blue:2003nk} to drive plasma wakefields with accelerating gradients up to ${\sim}100\,\mathrm{ GV/m}$, roughly three orders of magnitude above conventional radiofrequency (RF) cavities. Experiments with intense lasers \cite{Picksley:2024cdd} and electron beams \cite{Blumenfeld:2007ph,Litos:2014yqa} have demonstrated multi-GeV acceleration at tens of GV/m. Because individual laser pulses and electron drivers carry limited total energy, reaching TeV-class energies with high beam intensity generally requires staging multiple plasma modules \cite{hogan2016electron,schroeder2010physics}. By contrast, the AWAKE experiment at CERN demonstrated proton-driven plasma wakefield acceleration \cite{Muggli:2016pou,Adli:2016rwp,Caldwell:2015rkk}, accelerating electrons to $2\,\mathrm{GeV}$ in a $10\,\mathrm{m}$ plasma driven by a high-intensity proton bunch. Because the proton bunch carries ${\sim}19\,\mathrm{kJ}$ of kinetic energy, this technique holds the potential to accelerate witness electrons to the TeV scale in a single acceleration stage \cite{AWAKE:2018gdq}.

\subsection{\label{sec:hi-driver}Heavy ion beams as drivers}

In our previous work \cite{li2025numericalinvestigationsheavyion}, we showed that heavy ion beams offer distinct advantages for plasma wakefield acceleration. For a given particle number and bunch geometry, their high charge state produces a high beam charge density capable of exciting strong wakefields, while their large ion mass and high magnetic rigidity help maintain wakefield structure over extended propagation distances. Heavy-ion bunches at modern facilities such as HIAF also carry substantial stored kinetic energy, offering potential for high-energy, high-intensity witness beams. However, due to their relatively low velocity, the dephasing length of the witness beam is short, which restricts single-stage energy gain. Therefore, the key challenge for heavy ion driven plasma wakefield acceleration is to extend the dephasing length.

Here, we investigate a drift-like phase-shift scheme that inserts periodic vacuum gaps between uniform plasma sections to rephase the witness electron bunch. Idealized multistage LCODE simulations show acceleration of a $16\,\mathrm{MeV}$ electron bunch to a peak energy of $700\,\mathrm{MeV}$ over $1.65\,\mathrm{m}$, and two-dimensional electromagnetic PIC simulations in EPOCH evaluate single-unit beam transport across the first plasma--vacuum--plasma unit under an external quadrupole lattice. This paper is organized as follows: Section~\ref{sec:dephasing} analyzes the dephasing dynamics, Section~\ref{sec:scheme} introduces the drift-like phase-shift acceleration scheme, and Section~\ref{sec:simulations} presents the LCODE and EPOCH simulation results. Conclusions are summarized in Section~\ref{sec:conclusion}.

\section{\label{sec:dephasing}Dephasing dynamics}

In this study, we adopt heavy-ion parameters based on the High Intensity Heavy-ion Accelerator Facility (HIAF) and its planned upgrade (HIAF-Upgrade, HIAF-U) in China \cite{Yang:2023hfx,sun2020huizhou}. Following its first beam commissioning, HIAF is expected to deliver high-energy, high-intensity heavy-ion beams suitable for experimental studies of heavy-ion-driven plasma wakefield acceleration. Because the bunch lengths from HIAF are typically several meters ($\sigma_z \gg \lambda_{p}$), a raw bunch cannot directly drive a high-amplitude plasma wake.

When such a long ion bunch enters a plasma, the weak initial wakefield generated by its head imposes alternating focusing and defocusing transverse forces on the tail at the plasma wavelength $\lambda_{p}$. This drives the self-modulation instability (SMI) \cite{Kumar:2010bc,AWAKE:2023ssy,Caldwell:2011ir,Schroeder:2011hkj}, splitting the long bunch into a resonant microbunch train. To ensure a stable modulation and suppress competing instabilities, the seeded self-modulation (SSM) scheme \cite{AWAKE:2020stp,AWAKE:2022kmf,Lotov:2012ck,AWAKE:2017ulm} is typically employed, using either a preceding bunch or a sharp leading-edge charge distribution to pre-generate the initial transverse wakefields. During SMI growth, the wake phase velocity drops below the driver velocity, making the wake unfavorable for stable electron acceleration. Once modulation saturates, the wakefield stabilizes. In this work, we assume that the bismuth beam has already reached saturation as a preformed microbunch train and that the witness electrons are already trapped.

\begin{figure}[h]
	\centering
	\includegraphics[width=8.6cm]{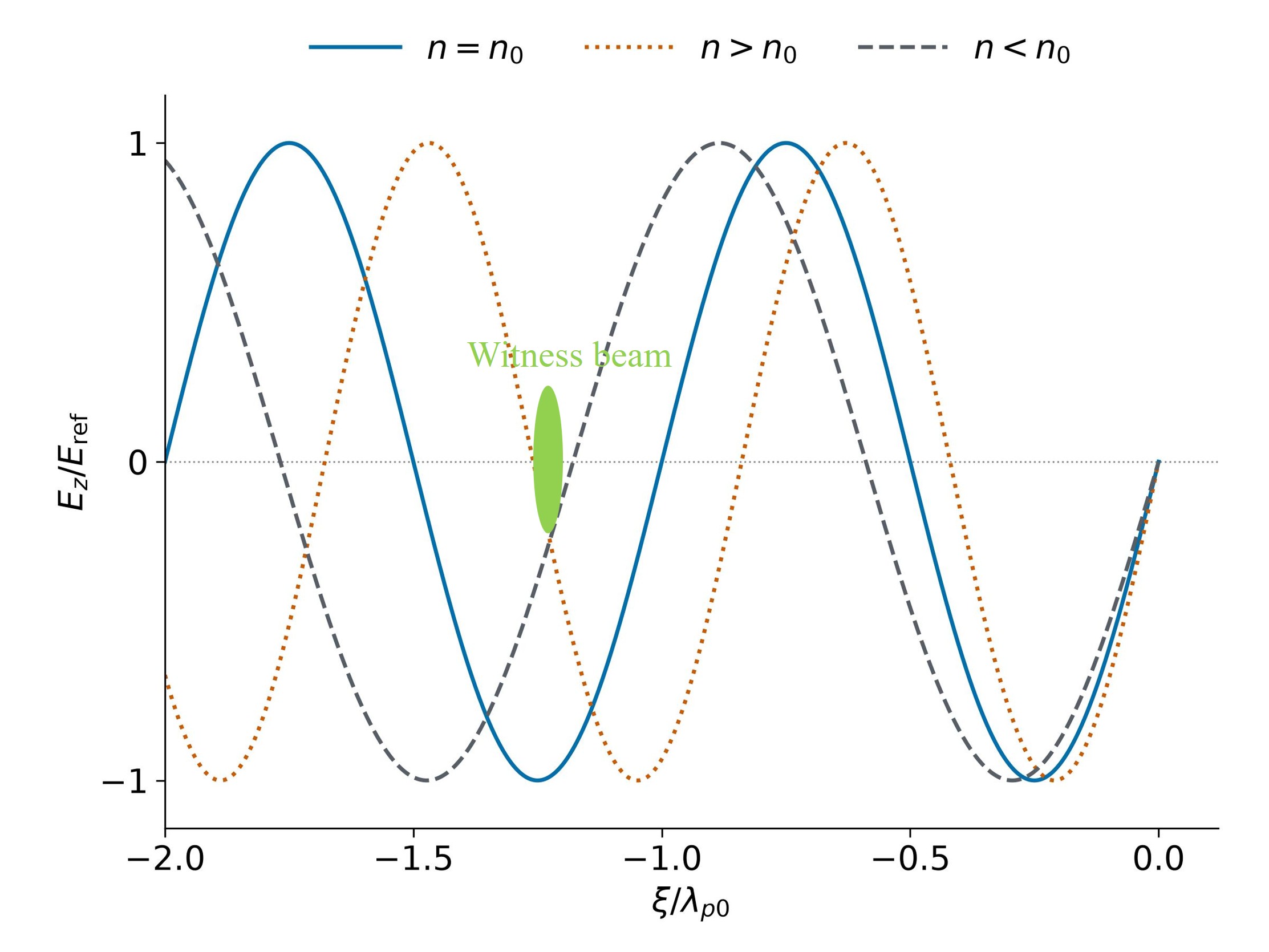}
	\caption{\label{fig:Figure_2}(Color) Phase of the accelerated electron bunch in plasmas of matched density $n = n_0$ (solid line), increased density $n > n_0$ (dotted line), and reduced density $n < n_0$ (dashed line), where $\lambda_{p0}$ is the plasma wavelength at $n_0$ \cite{li2025numericalinvestigationsheavyion}. Adapted from Li J, Xia G, Yang J, Liu J, Zhu R and Li G 2026 Phys. Plasmas 33 033102; doi: 10.1063/5.0316747; licensed under a Creative Commons Attribution (CC BY) license.}
\end{figure}

For a fully self-modulated driver beam in plasma wakefield acceleration, the dephasing length is determined by the relative velocity difference between the driver and the witness beam:
	
\begin{equation}\label{eq:dephasing_length}
	L_d = \frac{\lambda_{p} \beta_w}{4 \Delta \beta}
\end{equation}
where $\beta_d=v_d/c$ and $\beta_w=v_w/c$ are the normalized driver and witness velocities, $\Delta \beta = \beta_w - \beta_d$, and the factor of $4$ corresponds to the $\lambda_{p}/4$ phase window in a linear plasma wakefield that simultaneously provides longitudinal acceleration and transverse focusing. For a $^{209}\mathrm{Bi}^{83+}$ driver at a kinetic energy of $9.58\,\mathrm{ GeV/u}$ ($\gamma_d \approx 11.28$, $\beta_d \approx 0.9961$) and a $16\,\mathrm{ MeV}$ witness electron bunch ($\beta_w \approx 0.9995$) in a plasma of density $n_{p} = 2.8 \times 10^{15}\,\mathrm{ cm}^{-3}$ ($\lambda_{p} \approx 0.628\,\mathrm{ mm}$), the initial dephasing length is $L_d \approx 4.5\,\mathrm{cm}$ (decreasing slightly to ${\sim}4.0\,\mathrm{cm}$ as $\beta_w \to 1$). This short dephasing length intrinsically limits single-stage energy gain. The witness electrons therefore quickly outrun the accelerating and focusing phase and slip into deceleration, degrading beam quality.

To optimize the electron acceleration process, plasma density modulation is a commonly adopted method \cite{Petrenko:2015cxx,Braunmuller:2020aqw,AWAKE:2021vyl}. Once self-modulation has saturated and the witness is trapped, changing the plasma density $n_{p}$ primarily shifts the wake phase via the plasma wavelength,

\begin{align}\label{eq:plasma_wavelength}
	\lambda_{p} = \frac{2 \pi c}{\omega_{p}} = 2 \pi c \sqrt{\frac{\varepsilon_0 m_e}{n_{p} e^2}}
\end{align}
where $c$ is the speed of light in vacuum, $\omega_{p}$ is the electron plasma frequency, $\varepsilon_0$ is the vacuum permittivity, and $m_e$ is the electron rest mass. As illustrated in Fig.~\ref{fig:Figure_2}, raising the plasma density shortens $\lambda_{p}$ and shifts the wake forward relative to the witness; this can keep electrons in strong accelerating fields but risks pushing them into a defocusing region and causing beam loss. Lowering the density lengthens $\lambda_{p}$, moving the witness toward weaker acceleration or deceleration \cite{Lotov:2012ce,Katsouleas:1986zz}. 

\begin{figure}[h]
	\centering
	\begin{subfigure}{0.45\textwidth}
		\centering
		\includegraphics[width=\linewidth]{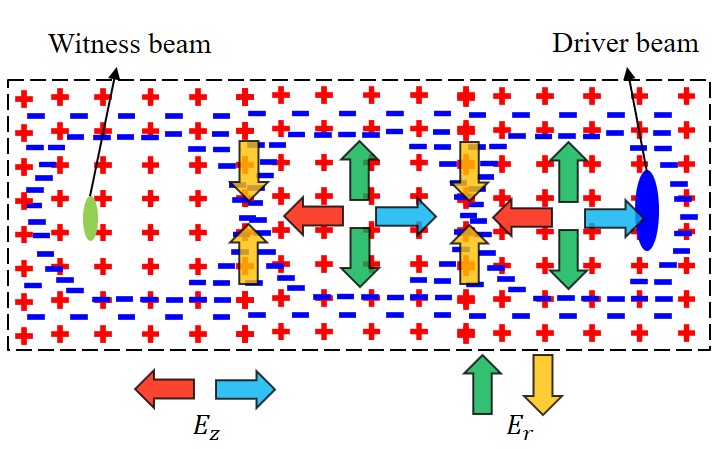}
		\subcaption{}
		\label{fig:Figure_3a}
	\end{subfigure}
	\vfill
	\begin{subfigure}{0.45\textwidth}
		\centering
		\includegraphics[width=\linewidth]{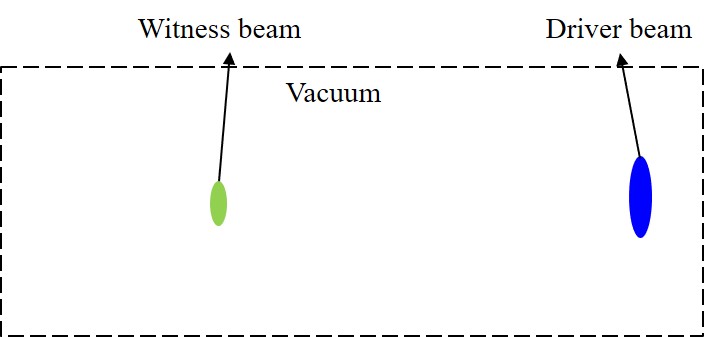}
		\subcaption{}
		\label{fig:Figure_3b}
	\end{subfigure}
	\vfill
	\begin{subfigure}{0.45\textwidth}
		\centering
		\includegraphics[width=\linewidth]{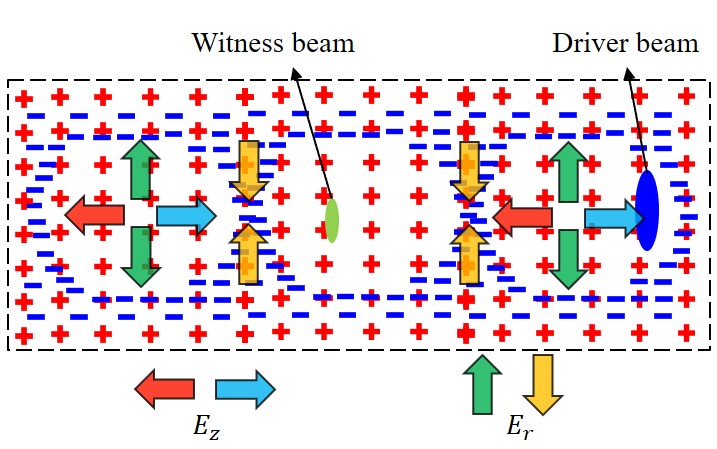}
		\subcaption{}
		\label{fig:Figure_3c}
	\end{subfigure}
	\caption{(Color) Principle of the drift-like phase-shift acceleration scheme. (a) Initial injection of the witness bunch into the accelerating and focusing phase; (b) vacuum drift introduced before the witness enters the decelerating phase; (c) restoration of the plasma density once the witness reaches the accelerating and focusing phase of the next wake bucket.}
	\label{fig:Figure_3}
\end{figure}

A common approach to mitigating the dephasing effect is to introduce upward plasma density ramps or steps, which continuously shorten the plasma wavelength. In this way, when the witness beam approaches the dephasing point, the local wave phase is shifted forward to keep the electrons within the accelerating and focusing region. However, such a continuous density increase also disrupts the match between the driver beam's rms radius and the plasma skin depth. As a result, the wakefield excited by the driver beam gradually weakens and eventually vanishes, thereby limiting the energy gain of the witness beam. This demonstrates that simply increasing the plasma density is insufficient to fully overcome dephasing, requiring an alternative phase-shift mechanism.

\section{\label{sec:scheme}Drift-like phase-shift acceleration scheme}

The proposed scheme draws on the concept of virtual drift sections in laser wakefield acceleration \cite{sadler2020overcoming}, where local plasma density steps suppress wake excitation so that electrons can slip forward in a nearly field-free region. Adapting this idea to heavy-ion-driven wakefields, we replace tailored density steps with vacuum gaps between uniform plasma sections. Inside each vacuum gap, the plasma wake vanishes completely. Because the heavy-ion driver remains subluminal ($v_d < c$) while the witness electrons are ultra-relativistic ($v_w \approx c$), the witness advances relative to the ion train during vacuum propagation, bypassing the decelerating and defocusing wake phases before re-entering the next plasma stage.

As illustrated in Fig.~\ref{fig:Figure_3}, the principle of the drift-like phase-shift scheme consists of three stages:
\begin{enumerate}
\item \textit{Acceleration} [Fig.~\ref{fig:Figure_3a}]: The witness electron bunch is injected into the accelerating and focusing phase of the plasma wakefield. Here, electrons gain energy from the wakefield.
\item \textit{Vacuum drift} [Fig.~\ref{fig:Figure_3b}]: The first plasma section ends before the witness enters a strongly decelerating phase. The witness then traverses a prescribed vacuum gap and advances relative to the driver to shift its longitudinal phase.
\item \textit{Re-entry and reacceleration} [Fig.~\ref{fig:Figure_3c}]: In the subsequent uniform plasma stage, the heavy-ion driver re-excites the wakefield, and the electrons are positioned in the accelerating and focusing phase, enabling further energy gain.
\end{enumerate}

Compared to density-ramp approaches, phase control here is set geometrically via the vacuum-gap length while keeping every plasma section at the same matched density. This preserves the resonance condition between driver and plasma across successive stages and allows witness electrons to advance sequentially through the wake buckets toward the driver head.

\begin{figure}[h!]
	\centering
	\begin{subfigure}{0.95\linewidth}
		\centering
		\includegraphics[width=\linewidth]{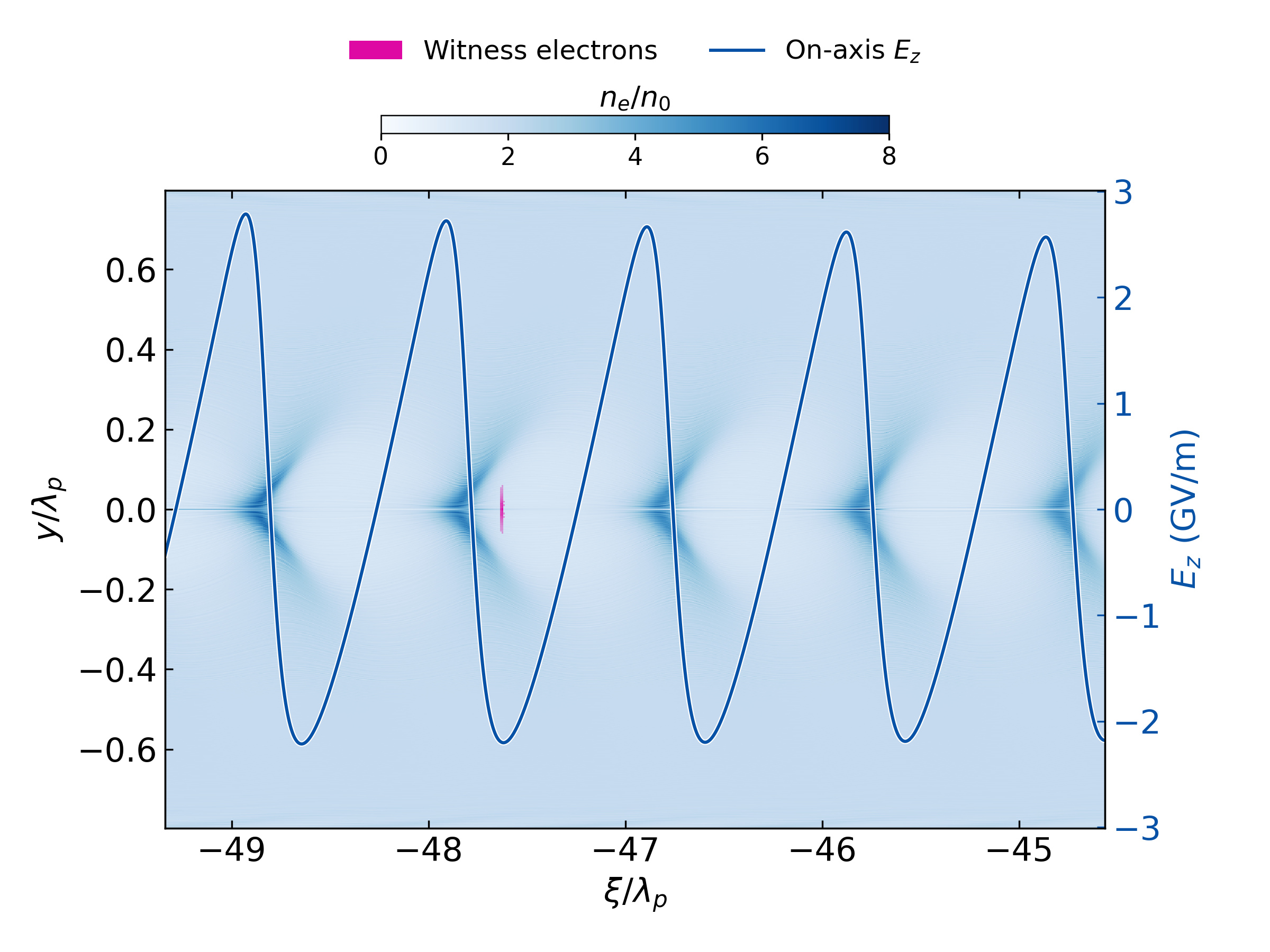}
		\subcaption{}
		\label{fig:Figure_4a}
	\end{subfigure}
	\vfill
	\begin{subfigure}{0.85\linewidth}
		\centering
		\includegraphics[width=\linewidth]{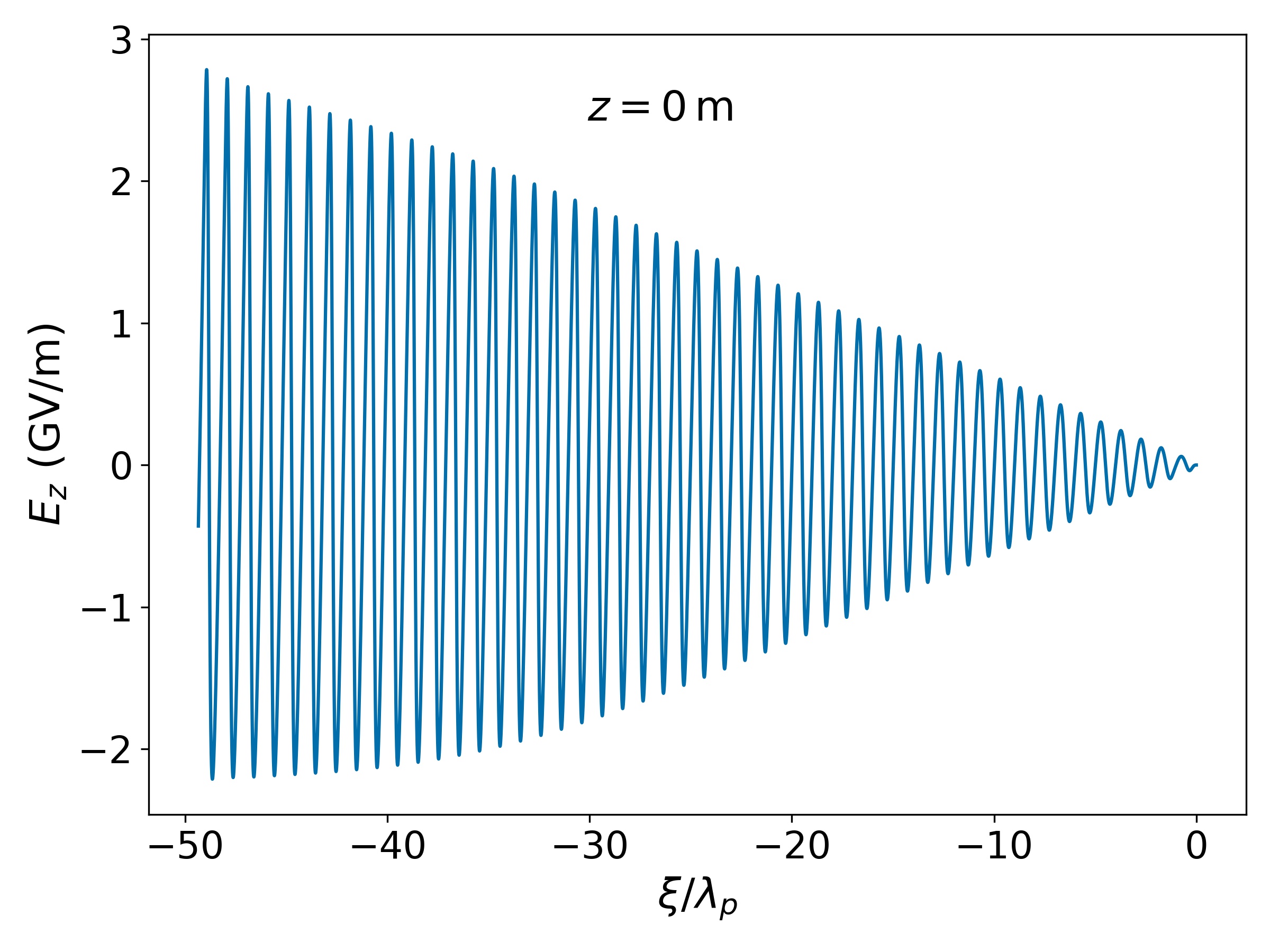}
		\subcaption{}
		\label{fig:Figure_4b}
	\end{subfigure}
	\caption{(Color) Initial beam and wakefield configuration in LCODE. (a) Normalized plasma electron density $n_e/n_0$ (color map), initial witness electron bunch (magenta), and on-axis longitudinal electric field $E_z$ (blue curve) near the tail of the 50-microbunch train. (b) On-axis longitudinal wakefield $E_z$ at $z=0\,\mathrm{m}$ versus normalized co-moving coordinate $\xi/\lambda_p$ ($\xi = z - ct$).}
	\label{fig:Figure_4}
\end{figure}

\begin{table}
\caption{\label{tab:elec_acc}The parameters of $^{209}\mathrm{Bi}^{83+}$ driver, witness electron bunch, plasma, and numerical setup used in the multistage LCODE simulations \cite{li2025numericalinvestigationsheavyion}.}
\begin{ruledtabular}
\begin{tabular}{lc}
Parameters &HIAF \\
\hline
Driver beam & $^{209}\mathrm{Bi}^{83+}$\\
Energy (GeV/u)& 9.58 \\
Peak number density ($\mathrm{cm}^{-3}$) & $1.26\times10^{12}$ \\
Microbunch radius (mm)& 0.1\\
Microbunch length (mm)& 0.314\\
Relative energy spread (\%)& 0.035\\
Plasma density ($\mathrm{cm}^{-3}$)& $2.8 \times 10^{15}$\\
\hline
Witness beam& electron\\
Energy (MeV)& 16 \\
Bunch radius (mm)& 0.01\\
Bunch length (mm)& 0.01\\
Relative energy spread (\%)& 0.035\\
\hline
Simulation grid step (mm)& 0.001\\
Simulation window size (mm)& 31\\
Particles in layer & 1000\\
Foc-period & $1 \times 10^{6}$\\
Foc-strength & 0.03\\
\end{tabular}
\end{ruledtabular}
\end{table}

\begin{figure}[h!]
	\centering
	\begin{subfigure}{0.85\linewidth}
		\centering
		\includegraphics[width=\linewidth]{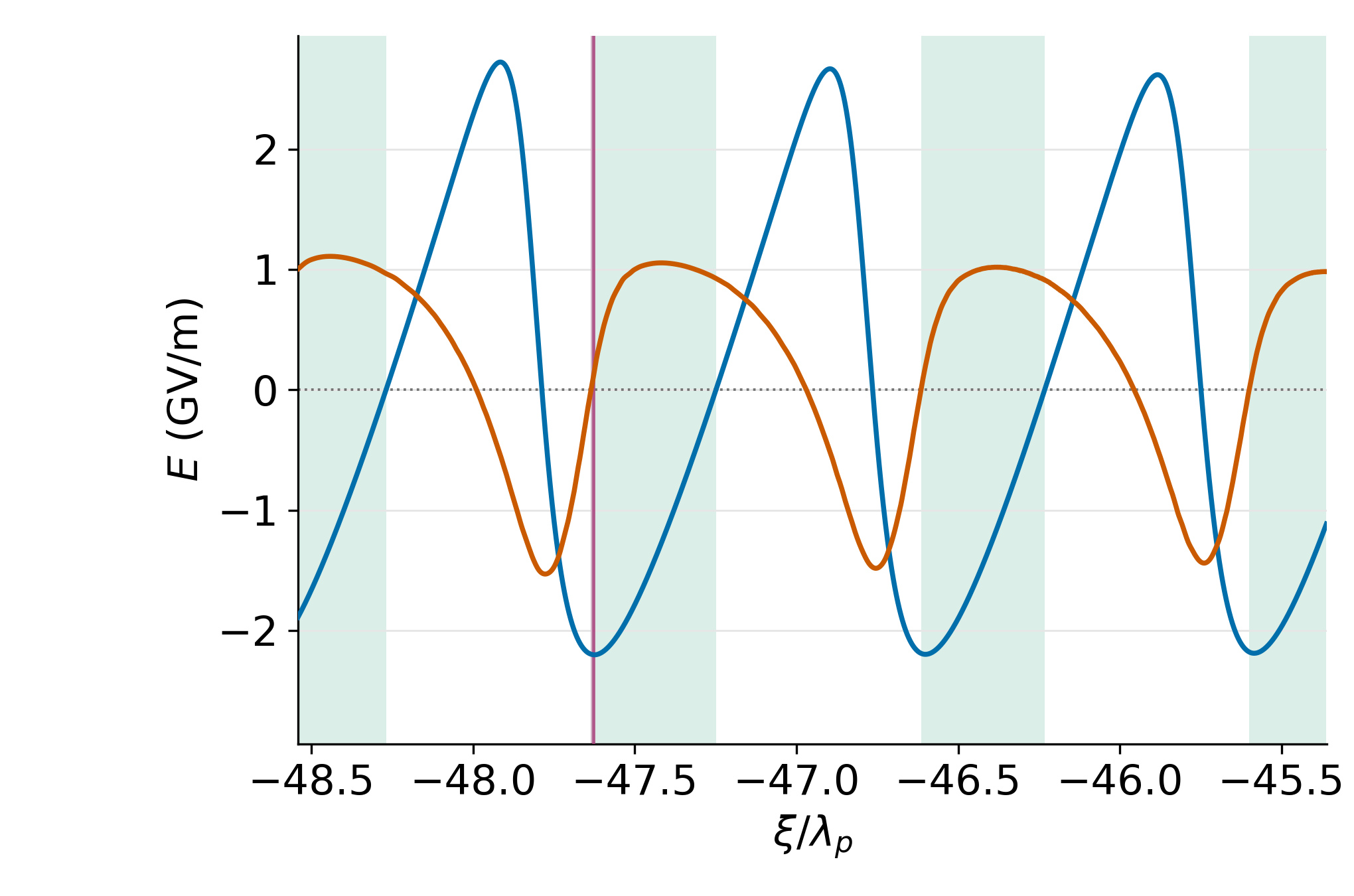}
		\subcaption{}
		\label{fig:Figure_5a}
	\end{subfigure}
	\vfill
	\begin{subfigure}{0.85\linewidth}
		\centering
		\includegraphics[width=\linewidth]{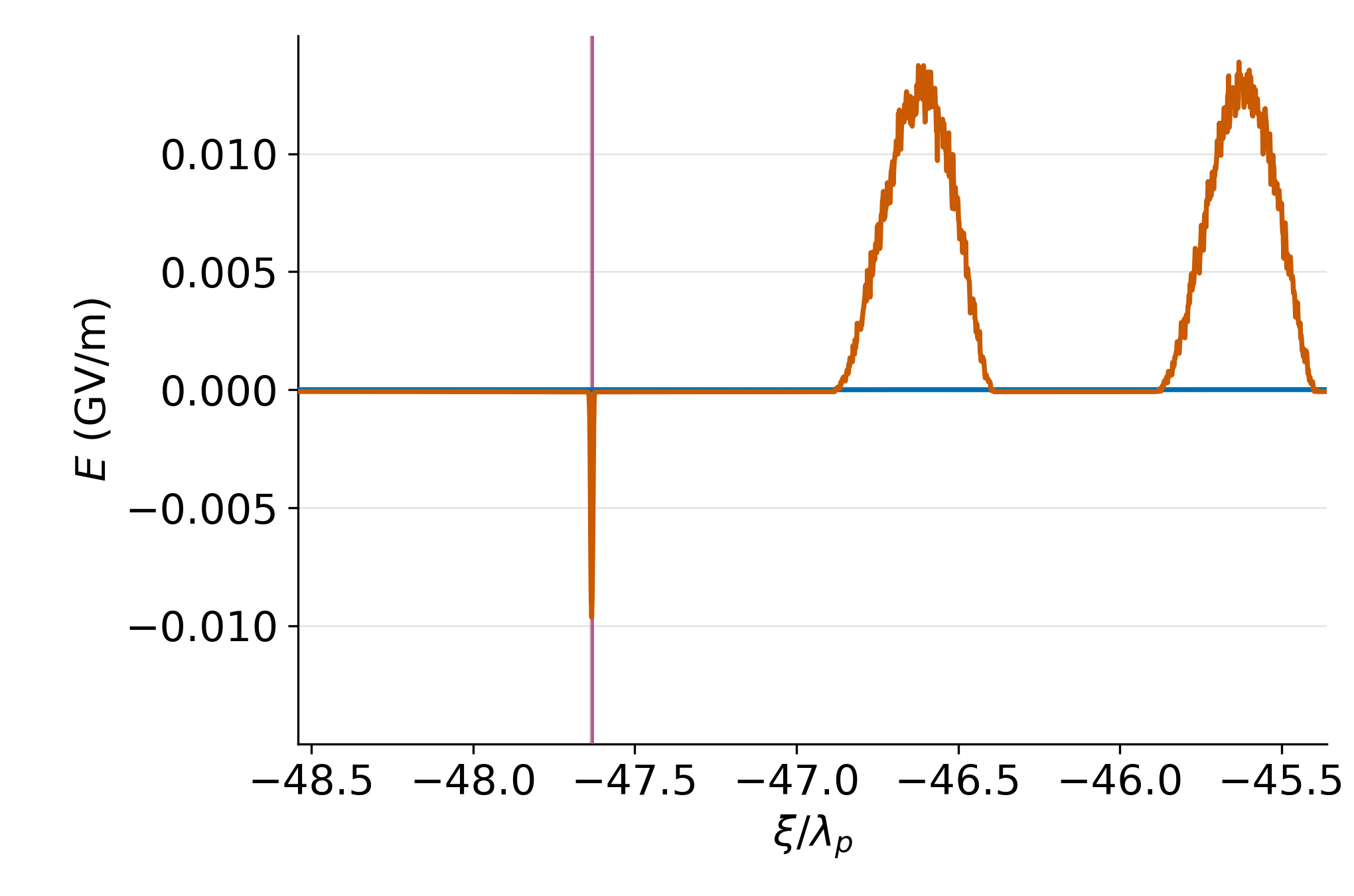}
		\subcaption{}
		\label{fig:Figure_5b}
	\end{subfigure}
	\vfill
	\begin{subfigure}{0.85\linewidth}
		\centering
		\includegraphics[width=\linewidth]{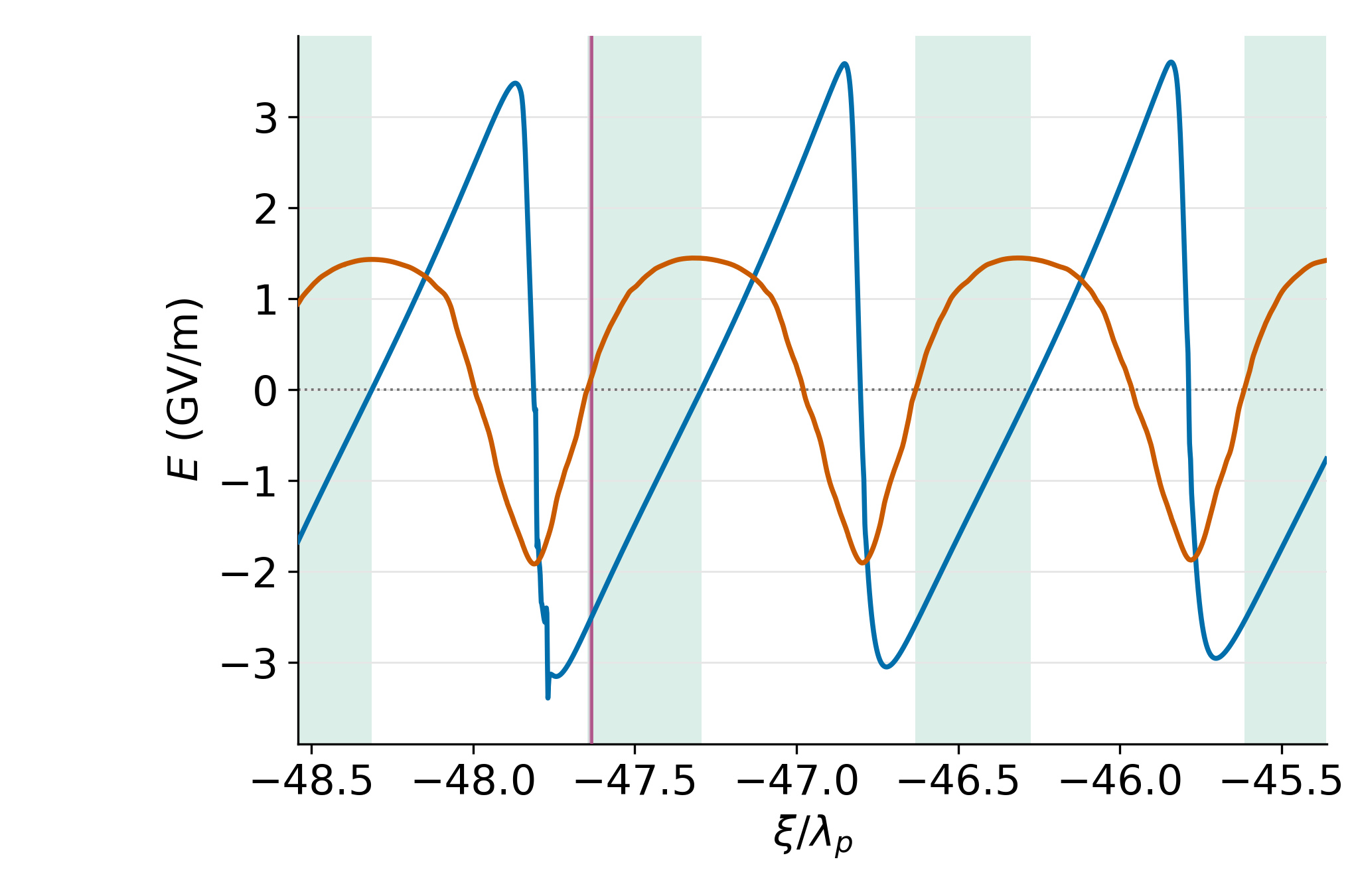}
		\subcaption{}
		\label{fig:Figure_5c}
	\end{subfigure}
	\caption{(Color) Evolution of the witness electron bunch and plasma wakefield in LCODE: (a) initial state ($z=0$), (b) snapshot inside the vacuum gap at $z = 0.06\,\mathrm{m}$, and (c) snapshot upon plasma restoration at $z = 0.14\,\mathrm{m}$. Blue and orange curves show the on-axis longitudinal field and near-axis transverse focusing field, respectively; green shading marks the accelerating and focusing phase region, and the purple vertical line marks the longitudinal position of the witness bunch. Note the reduced vertical scale in (b), where the plasma wake vanishes and only weak direct beam fields remain.}
	\label{fig:Figure_5}
\end{figure}

\section{\label{sec:simulations}Simulations for electron acceleration}

Our numerical study focuses on two complementary questions. First, quasi-static axisymmetric simulations with LCODE \cite{Sosedkin:2016} examine multistage energy gain under idealized transverse confinement. Second, two-dimensional electromagnetic PIC simulations with EPOCH evaluate the first plasma--vacuum--plasma unit with finite-length quadrupole fields, testing driver-plasma evolution, wake recovery, and witness transport.

\subsection{Idealized multistage simulations in LCODE}

LCODE operates in 2D3V axisymmetric geometry using a light-speed co-moving window ($\xi = z - ct$) and the quasi-static approximation. Under the quasi-static approximation, the beam is treated as rigid during the calculation of the plasma response. The key simulation parameters are listed in Table~\ref{tab:elec_acc}. To isolate longitudinal rephasing dynamics, we adopt a cold-beam approximation (zero initial emittance) \cite{saberi2024elevating}. Due to computational resource limitations, the simulations assume that the bismuth beam has already completed the self-modulation process and formed a stable train of microbunches.

To maintain transverse confinement across repeated vacuum gaps in the idealized LCODE model, a linear radial focusing force with LCODE normalized parameters foc-period $= 1\times10^6$ (in units of $c/\omega_{p}$) and foc-strength $= 0.03$ is applied. Specifically, transverse momentum kicks in the $r$ direction are applied to the particles within LCODE to model the external focusing effect for the beam.

Figure~\ref{fig:Figure_4} shows the initial states of the beam and excited wakefields in the simulation of the drift-like phase-shift acceleration scheme, where the horizontal axis is normalized as $\xi/\lambda_p$ ($\xi = z - ct$). Fig.~\ref{fig:Figure_4a} shows the initial witness bunch (magenta) positioned in the wake bucket alongside the normalized plasma electron density $n_e/n_0$ and on-axis longitudinal field $E_z$. Fig.~\ref{fig:Figure_4b} shows the longitudinal accelerating field generated by the 50 bismuth microbunches, reaching a peak amplitude of $2.2\,\mathrm{GV/m}$. The witness electrons start as a quasi-monoenergetic $16\,\mathrm{MeV}$ bunch. In the figure, $z = 0\,\mathrm{m}$ represents the initial position of the beam at the entrance of the plasma.

\begin{figure}[htbp]
	\centering
	\begin{subfigure}{0.85\linewidth}
		\centering
		\includegraphics[width=\linewidth]{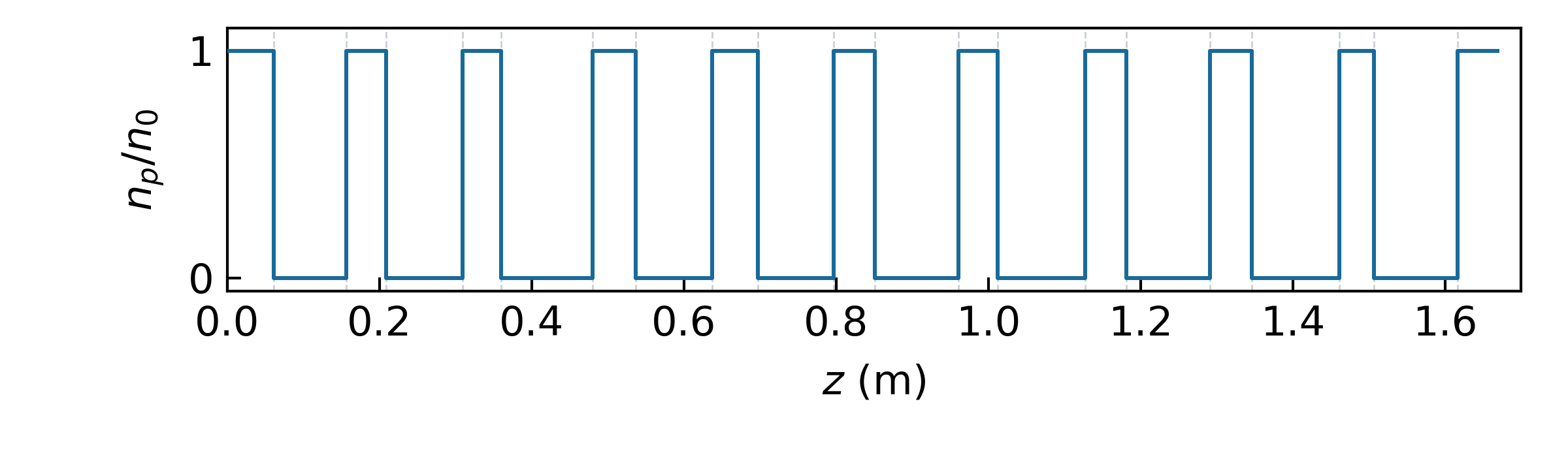}
		\subcaption{}
		\label{fig:Figure_6a}
	\end{subfigure}
	\vfill
	\begin{subfigure}{0.85\linewidth}
		\centering
		\includegraphics[width=\linewidth]{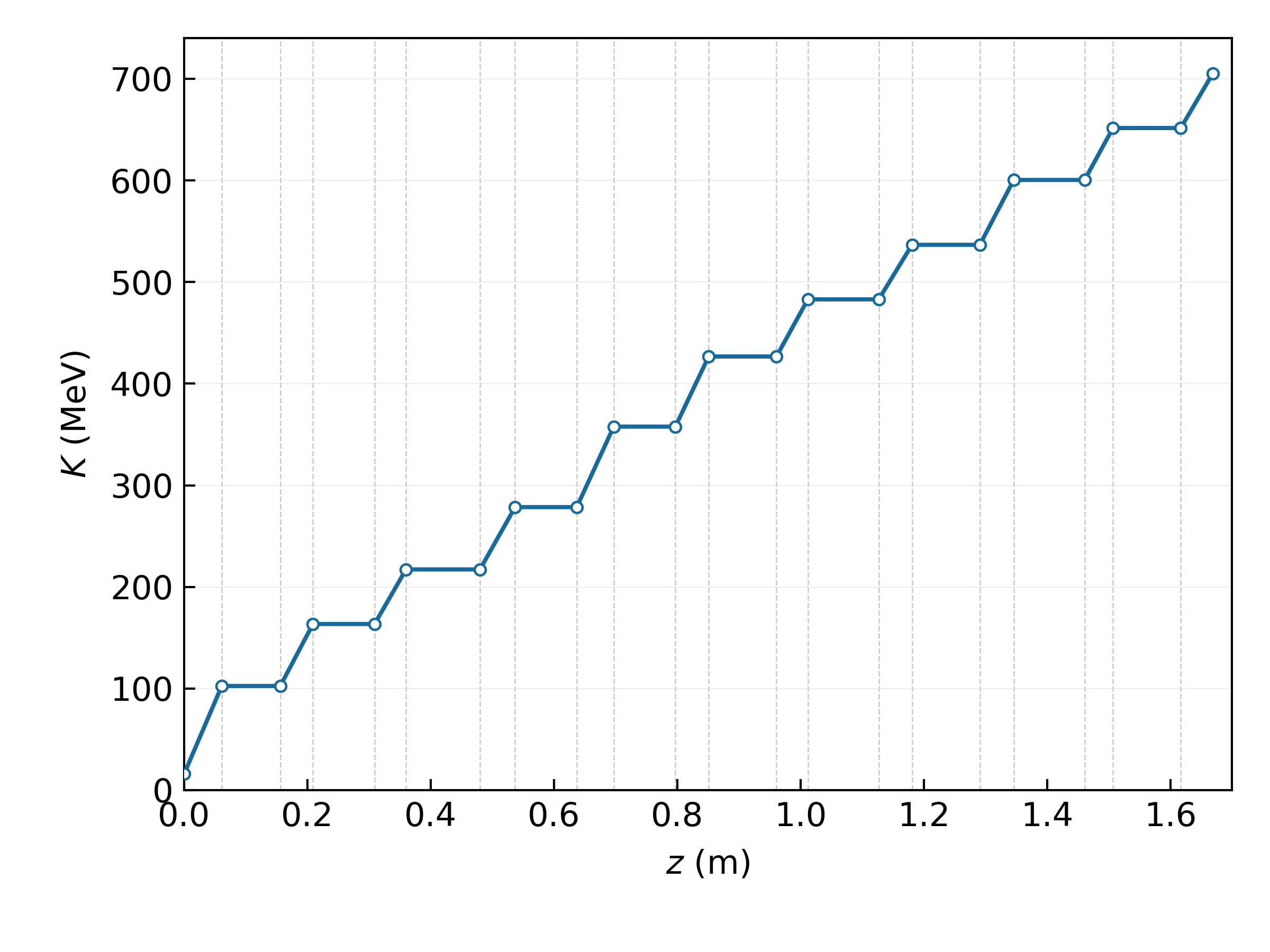}
		\subcaption{}
		\label{fig:Figure_6b}
	\end{subfigure}
	\vfill
	\begin{subfigure}{0.85\linewidth}
		\centering
		\includegraphics[width=\linewidth]{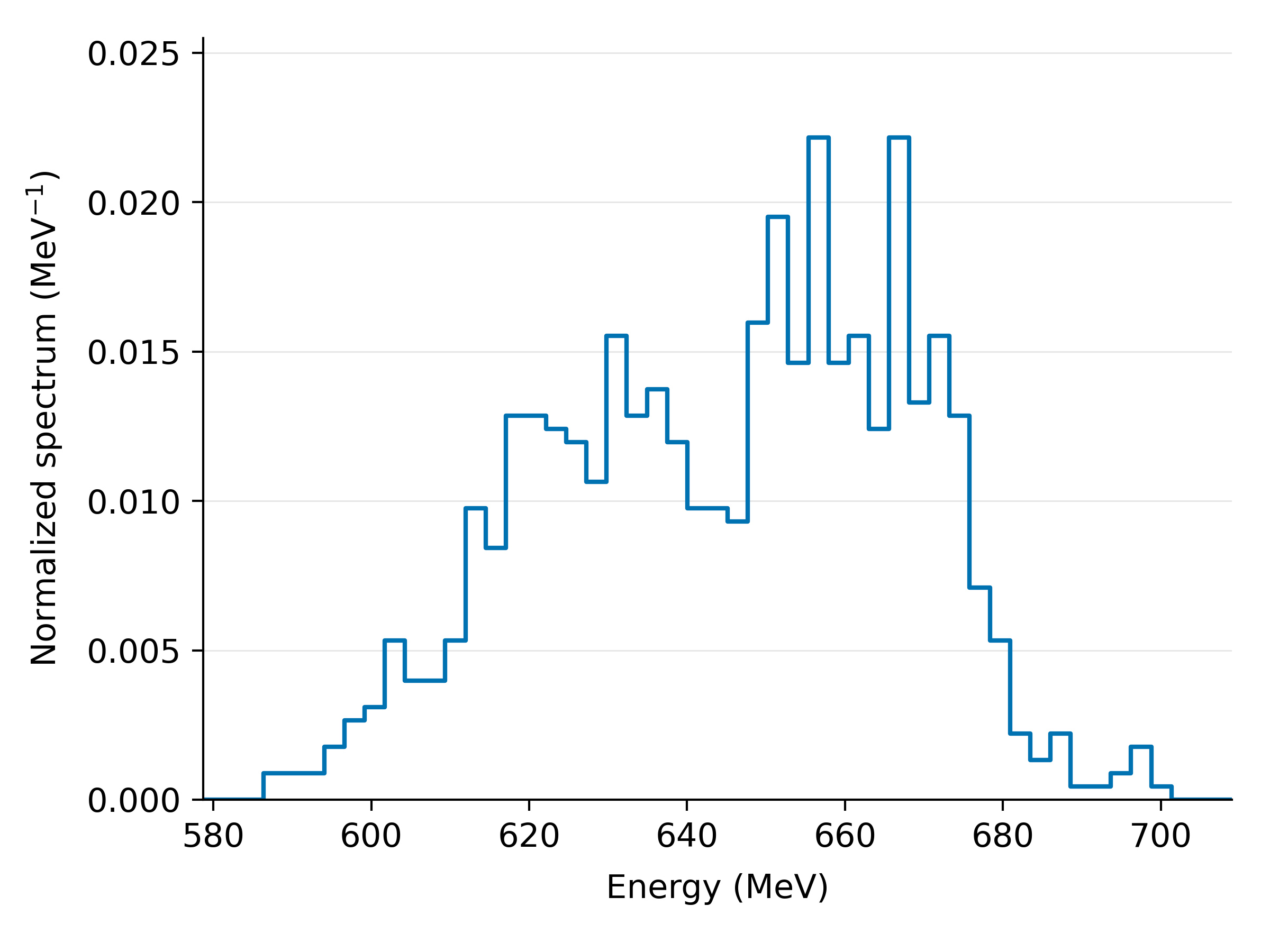}
		\subcaption{}
		\label{fig:Figure_6c}
	\end{subfigure}
	\caption{(Color) Multistage LCODE results: (a) normalized plasma density profile $n_p/n_0$ with $n_0 = 2.8 \times 10^{15}\,\mathrm{cm}^{-3}$, (b) evolution of the witness peak energy along $z$, and (c) final witness energy spectrum at $z = 1.65\,\mathrm{m}$.}
	\label{fig:Figure_6}
\end{figure}

The simulation process of plasma wakefield acceleration using the drift-like phase-shift scheme is illustrated in Fig.~\ref{fig:Figure_5}. Initially positioned in the accelerating and focusing phase [Fig.~\ref{fig:Figure_5a}], the witness electrons gain energy and slip forward toward the decelerating phase. To prevent them from entering the decelerating or defocusing regions, a vacuum section is introduced in the simulation, where the plasma wakefield vanishes, as shown in Fig.~\ref{fig:Figure_5b}. Within this vacuum region, the electrons drift freely until their relative phase shifts enough to avoid the decelerating and defocusing regions of the subsequent plasma stage. When the electrons are about to enter the next plasma acceleration stage, the plasma density is restored to re-establish the wakefield, allowing the electrons to enter the accelerating and focusing region, as shown in Fig.~\ref{fig:Figure_5c}. By carefully designing the distribution of plasma and vacuum sections, the electrons can avoid the decelerating and defocusing regions throughout propagation, sustaining continuous acceleration over extended distances. In principle, this enables the electrons to be accelerated from the tail to the head of the heavy-ion driver beam, fully utilizing the excited plasma wakefield.

\begin{figure}[htbp]
	\centering
	\includegraphics[width=8.0cm]{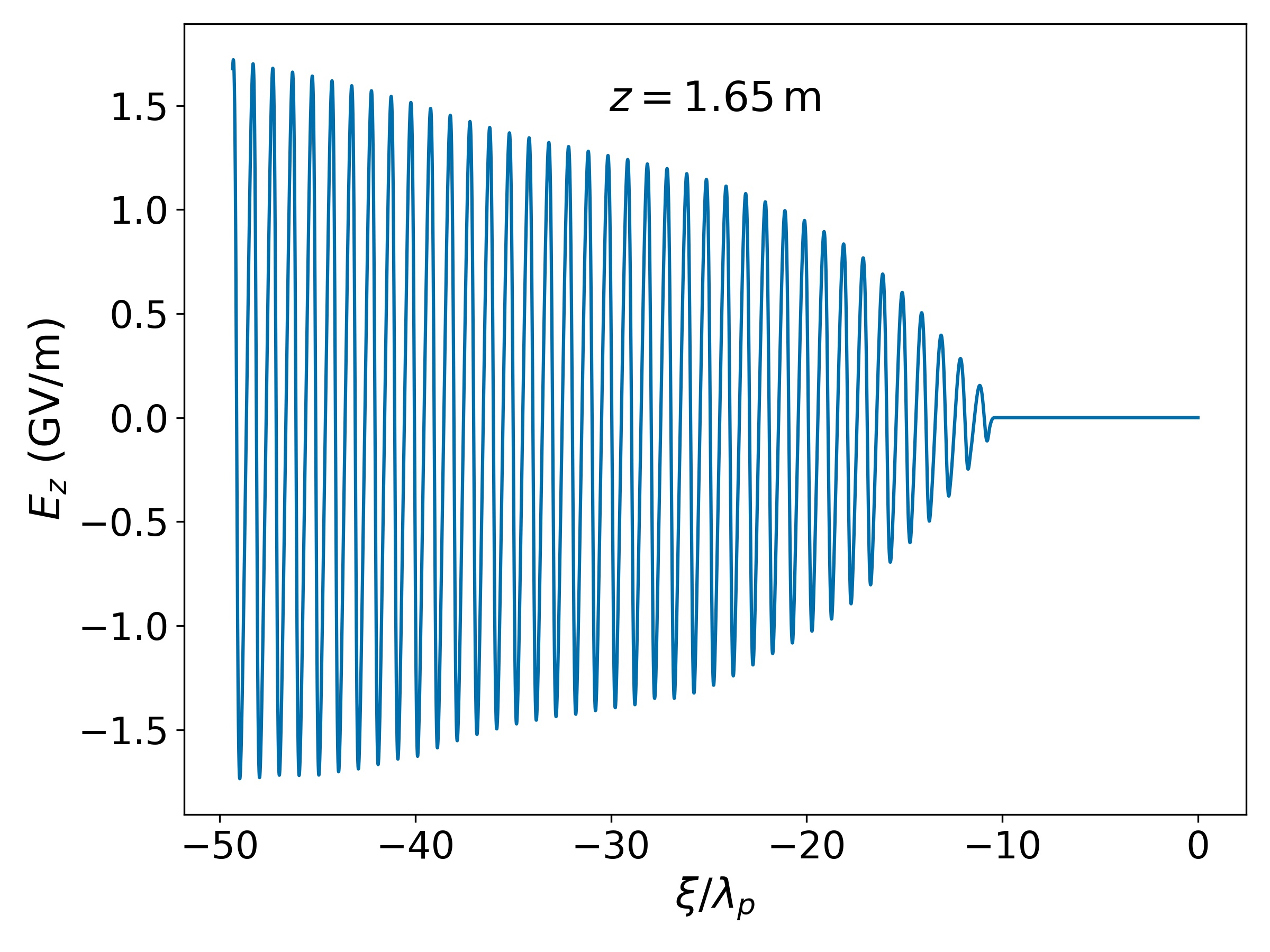}
	\caption{\label{fig:Figure_7}(Color) On-axis longitudinal plasma wakefield excited by the $^{209}\mathrm{Bi}^{83+}$ microbunch train at $z = 1.65\,\mathrm{m}$ in the LCODE simulation.}
\end{figure}

Figure~\ref{fig:Figure_6} presents the final simulation results. Fig.~\ref{fig:Figure_6a} shows the normalized plasma density profile $n_p/n_0$ and the initial plasma density is $ n_0 = 2.8 \times 10^{15}\,\mathrm{cm}^{-3}$. In the first $5.6\,\mathrm{cm}$ of beam propagation, the plasma density remains constant. When the accelerated electrons approach the dephasing point, a vacuum section of $8.2\,\mathrm{cm}$ is introduced. Within this region, the electrons drift freely in vacuum until they re-enter the accelerating and focusing phase of the plasma wakefield, at which point the plasma density is restored and acceleration resumes. Phase tracking confirms that inserting each vacuum gap before dephasing keeps the electrons out of the decelerating and defocusing phases. Figure~\ref{fig:Figure_6b} shows the evolution of the electron peak energy along the propagation distance. As illustrated in Fig.~\ref{fig:Figure_6c}, with this alternating plasma--vacuum sequence, the electrons are accelerated from $16\,\mathrm{MeV}$ to a peak energy of $700\,\mathrm{MeV}$ over a total distance of $1.65\,\mathrm{m}$, with an energy spread of about $5\%$. The corresponding effective accelerating gradient of the plasma wakefield reaches approximately $415\,\mathrm{MV/m}$.

Figure~\ref{fig:Figure_7} shows the longitudinal wakefield at $z = 1.65\,\mathrm{m}$. Compared with Fig.~\ref{fig:Figure_4b}, the peak amplitude decreases from ${\approx}2.2$ to ${\approx}1.7\,\mathrm{GV/m}$. Because the subluminal $^{209}\mathrm{Bi}^{83+}$ driver ($\beta_d \approx 0.9961$) slips backward in the light-speed simulation window, the front region near $\xi = 0$ empties as the driver head recedes. Over $1.65\,\mathrm{m}$, this cumulative backward slip reaches $(1 - \beta_d)z \approx 6.5\,\mathrm{mm}$, meaning that the ultra-relativistic witness electrons advance ahead by roughly ten microbunch buckets toward the head of the finite $50$-microbunch train. Consequently, fewer upstream microbunches contribute coherently ahead of the witness position, accounting for both the lower wakefield amplitude in Fig.~\ref{fig:Figure_7} and the gradual reduction in per-stage energy gain in Fig.~\ref{fig:Figure_6b}. By keeping every plasma section at the same matched density, the drift-like phase-shift scheme mitigates dephasing while avoiding the mismatch between driver's size and plasma skin depth inherent to plasma density ramps or steps.

\begin{table}
\caption{\label{tab:pic-quadrupoles}
Baseline quadrupole settings within the first phase-shift unit. Axial positions are measured from the entrance of the first plasma stage, and signed integrated gradients include end transitions.}
\begin{ruledtabular}
\begin{tabular}{lcc}
Axial interval (mm) & Body gradient (T/m)
& $\int G\,\mathrm{d}z$ (T) \\
\hline
0--20    & $+600$ & $+10.20$ \\
30--50   & $-600$ & $-10.20$ \\
60--80   & $+440$ & $+7.48$  \\
90--115  & $-540$ & $-11.88$ \\
125--145 & $+450$ & $+7.65$  \\
155--175 & $-360$ & $-6.12$  \\
185--205 & $+600$ & $+10.20$ \\
\end{tabular}
\end{ruledtabular}
\end{table}

\subsection{First-unit simulations in EPOCH}
\label{sec:pic-rephasing}

The preceding LCODE simulations examined repeated rephasing under idealized transverse confinement. Here, we use two-dimensional electromagnetic PIC simulations with EPOCH to examine whether the $\mathrm{Bi}$ driver can cross the vacuum gap in the first rephasing unit, re-excite a wake in the restored plasma, and accelerate the same witness bunch further.

The driver consists of ten preformed $^{209}\mathrm{Bi}^{83+}$ microbunches with a nominal transverse rms width of $100\,\mu\mathrm{m}$. The witness electron bunch has an initial kinetic energy of ${\sim}16\,\mathrm{MeV}$, a transverse rms width of $10\,\mu\mathrm{m}$, and a full length of $10\,\mu\mathrm{m}$. In the baseline run, the witness electrons are tracked as passive particles. Along the propagation axis, the first plasma section, vacuum gap, and wake-recovery plasma section extend over $48.5$, $96.7$, and $60.0\,\mathrm{mm}$, respectively. Both plasma sections are initialized with a uniform density of $n_0 \approx 2.8\times10^{21}\,\mathrm{m}^{-3}$ ($\lambda_p = 2\pi c/\omega_p = 0.6283\,\mathrm{mm}$), while the vacuum gap contains only beam-generated electromagnetic fields and external magnetic fields.

\begin{figure}[htbp]
    \centering
    \includegraphics[width=0.85\linewidth]{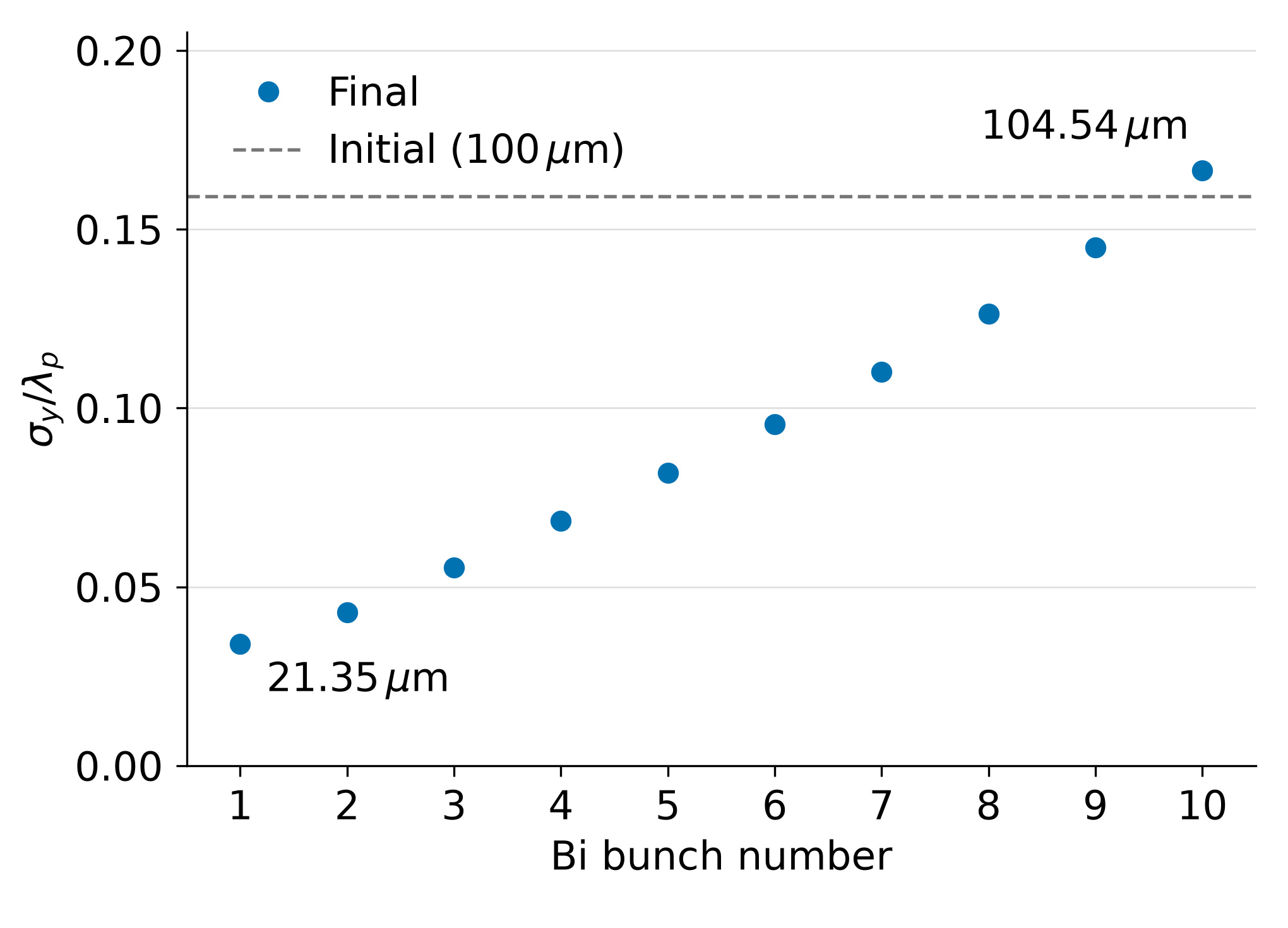}
    \caption{Final transverse rms widths of the ten $\mathrm{Bi}$ microbunches evaluated at the end of the first phase-shift unit for the converging baseline lattice. Values are normalized to $\lambda_p=0.6283\,\mathrm{mm}$, with the dashed line indicating the nominal initial width of $100\,\mu\mathrm{m}$.}
    \label{fig:Figure_8}
\end{figure}

External focusing is modeled using alternating-polarity linear quadrupole fields with finite magnet lengths, inter-magnet gaps, and fringe-field transitions. Through iterative PIC parameter scans at fixed plasma and vacuum lengths, we selected a baseline lattice (Table~\ref{tab:pic-quadrupoles}) capped at a peak gradient of $600\,\mathrm{T/m}$, with a $+450\,\mathrm{T/m}$ magnet located at $z=125$--$145\,\mathrm{mm}$. Positive and negative gradients focus and defocus the positively charged $\mathrm{Bi}$ ions in $y$, respectively, with the opposite polarity acting on the electrons. The $\mathrm{Bi}$ driver is initialized with $p_y/p_z=-\kappa_{\mathrm{Bi}}y$, where $\kappa_{\mathrm{Bi}}=3.5\,\mathrm{m}^{-1}$, corresponding to an initial rms convergence angle of about $0.35\,\mathrm{mrad}$.

Figure~\ref{fig:Figure_8} shows the final transverse rms widths of all ten $\mathrm{Bi}$ microbunches evaluated at the end of the simulation. At the end of the simulation, the rms widths of the first and tenth microbunches are $21.35$ and $104.54\,\mu\mathrm{m}$, respectively. The trailing tenth microbunch broadens by $4.54\%$ relative to its initial state, indicating limited transverse broadening of the bunch train across this finite-length unit.

\begin{figure}[htbp]
	\centering
	\begin{subfigure}{0.85\linewidth}
		\centering
		\includegraphics[width=\linewidth]{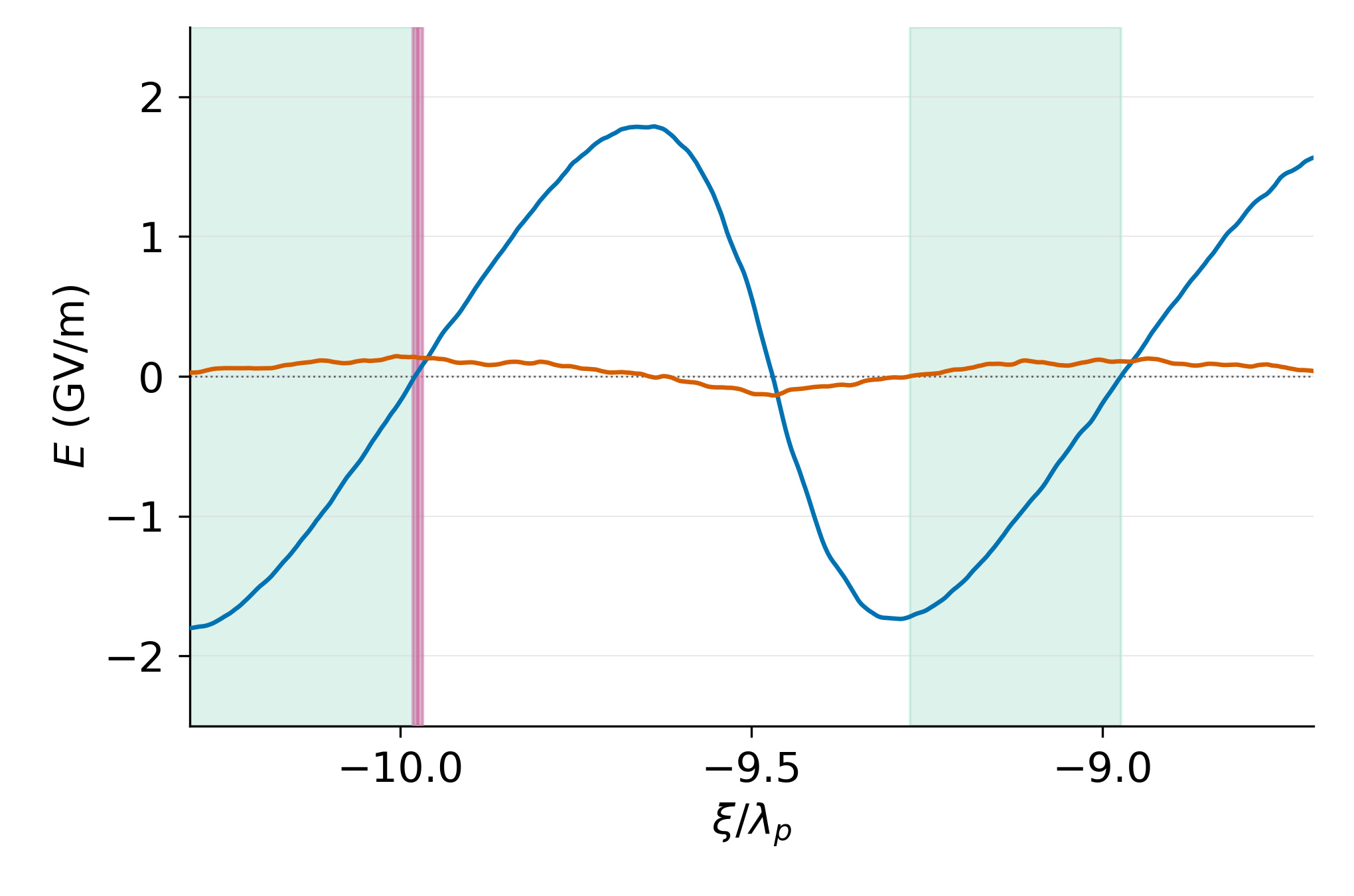}
		\subcaption{}
		\label{fig:Figure_9a}
	\end{subfigure}
	\vfill
	\begin{subfigure}{0.85\linewidth}
		\centering
		\includegraphics[width=\linewidth]{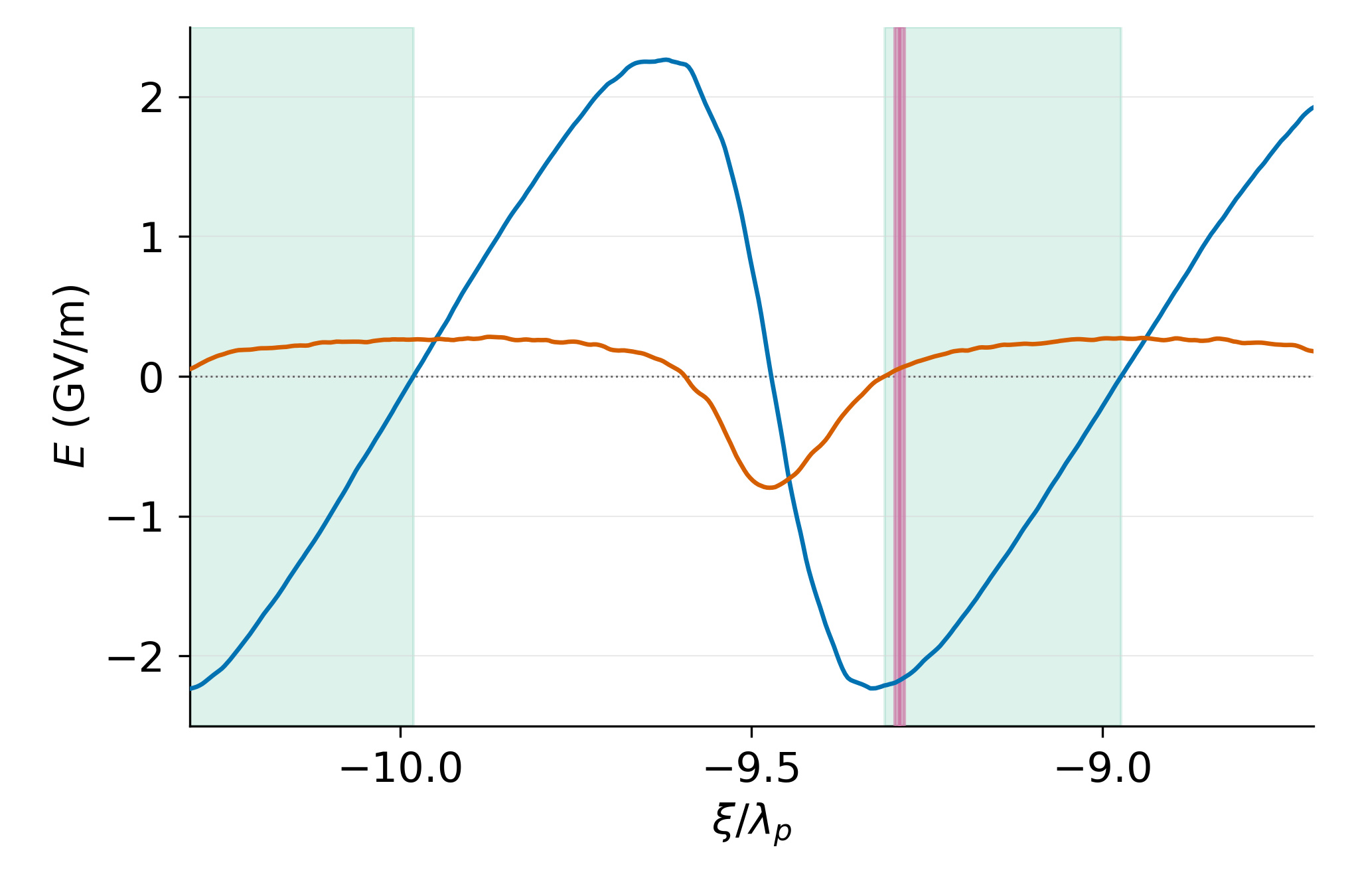}
		\subcaption{}
		\label{fig:Figure_9b}
	\end{subfigure}
	\vfill
	\begin{subfigure}{0.85\linewidth}
		\centering
		\includegraphics[width=\linewidth]{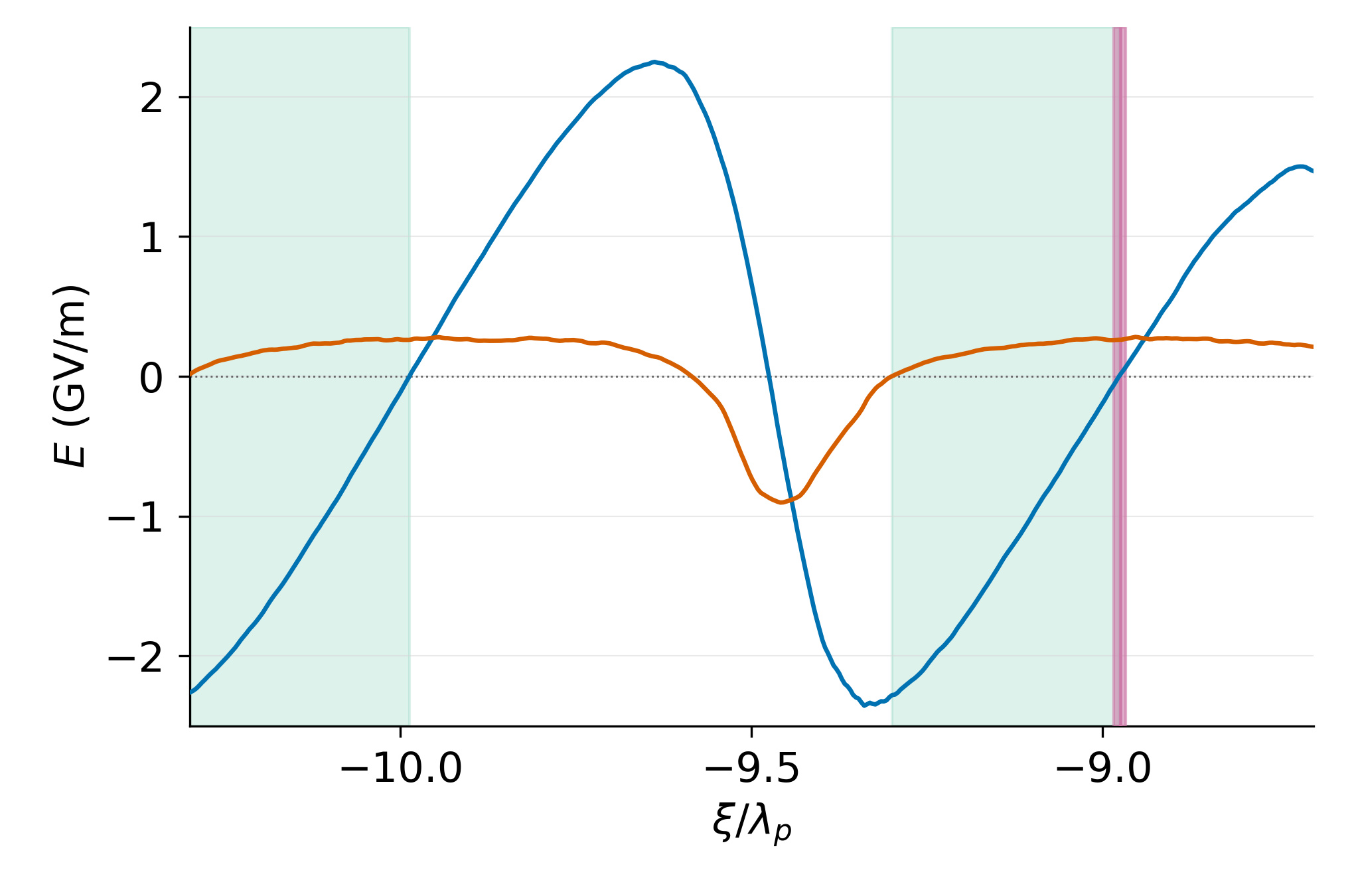}
		\subcaption{}
		\label{fig:Figure_9c}
	\end{subfigure}
	\caption{Phase relation between the converging witness bunch ($\kappa_e=100\,\mathrm{m}^{-1}$) and the wakefields in (a) the first plasma stage at an electron centroid position of $46\,\mathrm{mm}$, and in the restored plasma at depths of (b) $10\,\mathrm{mm}$ and (c) $60\,\mathrm{mm}$. Blue curves show the on-axis longitudinal field $E_z$, and orange curves show the transverse field $E_y$ near the axis. Green shading marks the accelerating and focusing region for electrons satisfying $E_z<0$ and $E_y > 0$, while the purple band and vertical line denote the longitudinal extent and centroid of the witness bunch.}
	\label{fig:Figure_9}
\end{figure}

Across the vacuum gap, the witness electrons slip forward by approximately $0.38\,\mathrm{mm}$ relative to the initial $\mathrm{Bi}$ driver. Figure~\ref{fig:Figure_9} tracks the longitudinal position of the converging witness bunch ($\kappa_e = 100\,\mathrm{m}^{-1}$) relative to the on-axis longitudinal electric field $E_z$ and the near-axis transverse focusing field $E_y$. Upon entering the second plasma stage, the electrons are placed inside the accelerating and focusing phase of the wake bucket ahead of their original one. The on-axis peak longitudinal field reaches $2.2\,\mathrm{GV/m}$ in the restored plasma, compared with $1.8\,\mathrm{GV/m}$ in the first stage, demonstrating effective wake re-excitation after vacuum transport.

\begin{table}
\caption{\label{tab:pic-witness}Passive witness electron parameters at the exit of the restored plasma section in the converging-driver baseline for two initial electron convergence parameters $\kappa_e$.}
\begin{ruledtabular}
\begin{tabular}{lccccc}
$\kappa_e$ & Energy & Energy spread & $\sigma_y$ & $\epsilon_{n,y}$ & Retention \\
($\mathrm{m}^{-1}$) & (MeV) & (\%) & ($\mu\mathrm{m}$) & ($\mathrm{mm\,mrad}$) & (\%) \\
\hline
0   & 136.15 & 8.19 & 45.10 & 130.57 & 96.84 \\
100 & 141.06 & 1.49 & 8.70  & 9.77   & 99.99 \\
\end{tabular}
\end{ruledtabular}
\end{table}

\begin{figure}[htbp]
    \centering
    \includegraphics[width=0.85\linewidth]{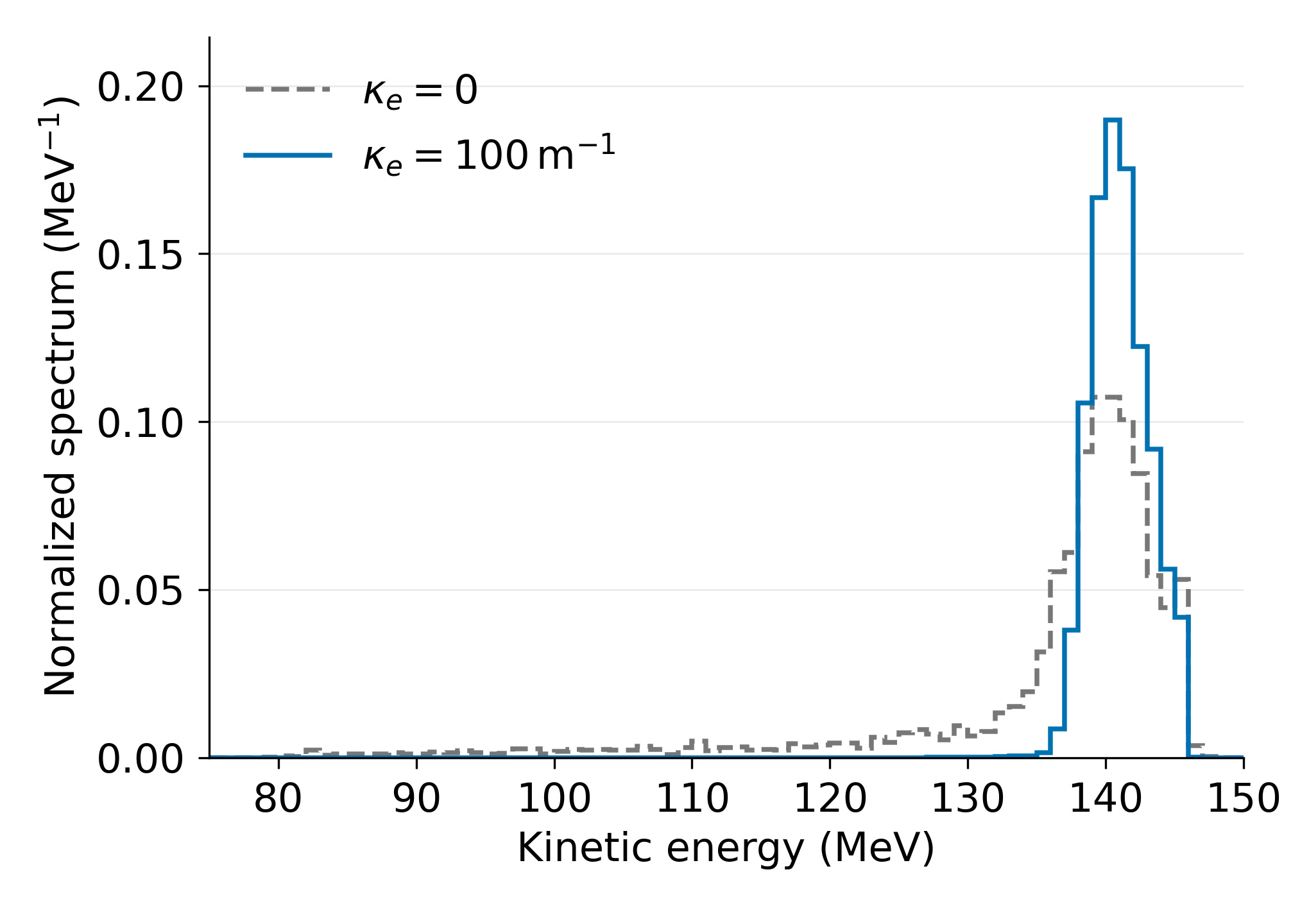}
    \caption{Normalized kinetic energy spectra of the two passive witness bunches at the exit of the restored plasma section in the converging driver baseline. The blue solid and gray dashed curves correspond to $\kappa_e=100\,\mathrm{m}^{-1}$ and $0$, respectively.}
    \label{fig:Figure_10}
\end{figure}

To assess the sensitivity to witness entrance optics, the baseline simulation simultaneously tracks an initially parallel witness bunch ($\kappa_e = 0$) and the converging bunch ($\kappa_e = 100\,\mathrm{m}^{-1}$) as passive test particles in the same electromagnetic fields. Their final kinetic energy spectra are shown in Fig.~\ref{fig:Figure_10}, and beam parameters are listed in Table~\ref{tab:pic-witness}. Compared with the parallel bunch, introducing initial convergence raises the mean kinetic energy from $136.15$ to $141.06\,\mathrm{MeV}$, reduces the relative rms energy spread from $8.19\%$ to $1.49\%$, and suppresses normalized emittance growth from $130.57$ to $9.77\,\mathrm{mm\,mrad}$ with $99.99\%$ macroparticle retention.

To test the tolerance to driver broadening without initial convergence, we performed a supplementary simulation with zero initial convergence for both beams ($\kappa_{\mathrm{Bi}} = 0$, $\kappa_e = 0$). Keeping the plasma sections and end quadrupoles unchanged, we replaced the three vacuum-gap magnets with a $-400\,\mathrm{T/m}$ quadrupole at $z=60$--$95\,\mathrm{mm}$ and a $+400\,\mathrm{T/m}$ quadrupole at $z=120$--$145\,\mathrm{mm}$ to rematch the initially parallel witness electrons. Without initial convergence, the tenth $\mathrm{Bi}$ microbunch broadens to an rms width of $139.12\,\mu\mathrm{m}$ at the end of the unit. With this ${\sim}39\%$ driver broadening, the on-axis longitudinal wakefield in the restored plasma still reaches a peak of $1.77\,\mathrm{GV/m}$, and the passive witness reaches a mean kinetic energy of $120.94\,\mathrm{MeV}$ with a $1.57\%$ energy spread, a normalized emittance of $6.93\,\mathrm{mm\,mrad}$, and $99.87\%$ retention. These results show that the first rephasing unit tolerates substantial trailing driver expansion without losing stable reacceleration or witness beam quality.

Finally, finite beam loading and witness self-fields were evaluated under the converging baseline conditions ($\kappa_{\mathrm{Bi}} = 3.5\,\mathrm{m}^{-1}$, $\kappa_e = 100\,\mathrm{m}^{-1}$) by enabling current deposition for the witness bunch at a peak number density of $5.96\times10^{20}\,\mathrm{m}^{-3}$, corresponding to an equivalent bunch charge of $0.3\,\mathrm{pC}$. Witness feedback slightly perturbs the driver or the restored wake: the final rms width of the tenth $\mathrm{Bi}$ microbunch shifts to $104.57\,\mu\mathrm{m}$ (within $0.03\%$ of the passive baseline), and the restored longitudinal wakefield amplitude changes by less than $0.24\%$. The witness bunch reaches a mean kinetic energy of $137.14\,\mathrm{MeV}$ with $99.99\%$ retention, while space-charge and beam-loading effects increase the relative energy spread to $3.10\%$, the rms width to $22.31\,\mu\mathrm{m}$, and the normalized emittance to $50.60\,\mathrm{mm\,mrad}$. Because the present PIC model is restricted to two dimensions with cold initial beams, evaluating finite initial emittance and full two-plane quadrupole transport remains necessary in future three-dimensional design work.

\section{\label{sec:conclusion}Conclusion}

Conventional rephasing methods based on longitudinal plasma density modulation face an intrinsic constraint in heavy-ion-driven plasma wakefield acceleration: increasing the plasma density mismatches the plasma skin depth relative to the driver size and shifts the microbunch resonance condition, which progressively degrades wakefield excitation. In this work, we investigated a drift-like phase-shift scheme that replaces plasma density modulation with periodic vacuum drift sections between uniform plasma stages. Using the velocity difference between ultra-relativistic witness electrons and subluminal $\mathrm{Bi}$ drivers, the witness slips across the decelerating and defocusing phases in vacuum before re-entering the accelerating bucket of the subsequent stage. Under idealized transverse confinement, multistage LCODE simulations indicate that a single $\mathrm{Bi}$ driver train can accelerate a $16\,\mathrm{MeV}$ electron bunch to a peak energy of $700\,\mathrm{MeV}$ over $1.65\,\mathrm{m}$ with ${\sim}5\%$ energy spread while maintaining wakefield amplitude across stages.

To examine beam transport beyond the quasi-static approximation, two-dimensional electromagnetic PIC simulations in EPOCH were performed for the first plasma--vacuum--plasma unit with finite-length alternating-gradient quadrupoles ($\le 600\,\mathrm{T/m}$). Under the prescribed converging entrance conditions ($\kappa_{\mathrm{Bi}} = 3.5\,\mathrm{m}^{-1}$, $\kappa_e = 100\,\mathrm{m}^{-1}$), broadening of the trailing tenth $\mathrm{Bi}$ microbunch is held to $4.54\%$, and the passive witness bunch is accelerated to $141.06\,\mathrm{MeV}$ with a $1.49\%$ relative energy spread and $9.77\,\mathrm{mm\,mrad}$ normalized emittance at the end of this unit. A supplementary zero-convergence configuration demonstrates that when the tenth $\mathrm{Bi}$ microbunch broadens to $139.12\,\mu\mathrm{m}$, wakefield recovery persists and the passive witness still reaches $120.94\,\mathrm{MeV}$ with a $1.57\%$ energy spread and $99.87\%$ retention. Enabling witness feedback at an equivalent charge of $0.3\,\mathrm{pC}$ under the converging baseline conditions preserves near-complete transmission ($99.99\%$) and yields $137.14\,\mathrm{MeV}$ with a $3.10\%$ energy spread, accompanied by moderate self-field-driven emittance growth.

The two-dimensional electromagnetic PIC results verify beam transport, wakefield recovery, and witness electron reacceleration across the first drift-like phase-shift unit under specified entrance conditions, providing a numerical basis for designing future experiments at facilities such as HIAF. Future three-dimensional studies will focus on multistage beam transport together with realistic beam emittance.

\section*{Acknowledgments}
	
This work was supported by the National Natural Science Foundation of China (No.~12425501 and No.~12605382).

\bibliography{aipsamp}

\end{document}